\documentclass[prd,twocolumn,superscriptaddress,floatfix,amsmath,amssymb]{revtex4-1}

\usepackage[normalem]{ulem}
\usepackage[english]{babel}
\usepackage{graphicx}
\usepackage{dcolumn}
\usepackage{bm}
\usepackage{verbatim}
\usepackage{mathrsfs}
\usepackage{cancel}
\usepackage{epstopdf}
\usepackage{color}

\newcommand{\ii}{\mathrm{i}}

\newcommand{\sgn}{\text{sgn}}

\usepackage[usenames,dvipsnames]{pstricks}
\usepackage{epsfig}
\usepackage{pst-grad} 
\usepackage{pst-plot} 

\usepackage{hyperref}

\usepackage{verbatim}

\begin{document}

\title{Anti-Unruh Phenomena}


\author{W. G. Brenna}
\email{wbrenna@uwaterloo.ca}
\affiliation{Department of Physics and Astronomy, University of Waterloo, Waterloo, Ontario N2L 3G1, Canada}

\author{Robert B. Mann}
\email{rbmann@uwaterloo.ca}
\affiliation{Department of Physics and Astronomy, University of Waterloo, Waterloo, Ontario N2L 3G1, Canada}

\author{Eduardo Mart\'{i}n-Mart\'{i}nez}
\email{emartinm@uwaterloo.ca}
\affiliation{Institute for Quantum Computing, University of Waterloo, Waterloo, Ontario, N2L 3G1, Canada}
\affiliation{Department of Applied Mathematics, University of Waterloo, Waterloo, Ontario, N2L 3G1, Canada}
\affiliation{Perimeter Institute for Theoretical Physics, Waterloo, Ontario, N2L 2Y5, Canada}

\begin{abstract}
We find that a uniformly accelerated particle detector coupled to the vacuum can cool down as its acceleration increases, due to relativistic effects. We show that in (1+1)-dimensions, a detector coupled to the scalar field vacuum for finite timescales (but long enough to satisfy the KMS condition) has a KMS temperature that decreases with acceleration, in certain regimes. This contrasts with the heating that one would expect from the Unruh effect. 
\end{abstract}

\maketitle

{\section{Introduction.} 
In 1976, it was proposed that the inequivalence of field quantization schemes associated with inertial and accelerated observers implied that observers uniformly accelerating in the Minkowski vacuum (as seen by inertial observers) would detect a thermal bath of particles \cite{Unruh1976}.  Specifically,  an accelerated particle detector coupled to the Minkowski vacuum would experience a thermal response \cite{Birrell1984}, a phenomenon known as the Unruh effect.  The temperature $T$ of this thermal bath was found to be proportional to the magnitude ${a}$ of the proper acceleration of the detector, with $T= {a}/2\pi$. The Unruh effect has been predicted and derived in contexts as disparate as axiomatic quantum field theory \cite{Sewell1982}, via Bogoliubov transformations \cite{Birrell1984}, and in studies of the response of non-inertial particle detectors both perturbatively \cite{Birrell1984} and non-perturbatively \cite{Brown2012,Bruschi2012,Hu:2012jr,Doukas:2013noa}, and even for non-uniformly accelerated trajectories \cite{Mann:2009dma,Ostapchuk:2011ud}.  More recently 
non-perturbative techniques developed in \cite{Brown2012} have been used to prove that within optical cavities in (1+1)-dimensions an accelerated detector equilibrates to a thermal state whose temperature is proportional to acceleration. This holds independently of the cavity boundary conditions, provided the detector is allowed enough interaction time \cite{Brenna2013}. 

Since all investigations so far have found that a particle detector coupled to the vacuum will detect more particles when it is accelerated than when undergoing inertial motion, we typically regard the Unruh effect as a universal phenomenon: simply put, `accelerated detectors get hotter'.
The common denominator in nearly all previous investigations is that the response of non-inertial detectors is studied for long interaction times, or for a field quantized in free infinite open space.  
However on empirical grounds, finite time studies with different boundary conditions are arguably relevant.
Any experimental setup based on quantum optics (e.g. an atom accelerating through an optical cavity) will necessarily require particular boundary conditions rather than infinite space.

But do accelerated detectors always become hotter?  In this paper we address this question using both perturbative and non-perturbative tools.  Previous numerical work on accelerating Unruh-deWitt detectors in cavities interacting for long times found that, as expected, a detector gets hotter and its temperature is proportional to its acceleration; $T\propto  {a}$ \cite{Brenna2013}. However, due to the finite length and time scales, the slope was not found to be $1/2\pi$. In this paper we find that when shorter interaction times comparable to the characteristic Heisenberg time of the detector are considered
the transition probability of an accelerated detector can actually {\it decrease} with acceleration. This is possible because even an inertial detector switched on for a finite time in the ground state, and coupled to the Minkowski vacuum, will not remain completely `cold' but will click due to switching noise and vacuum fluctuations (see \cite{Louko2008} and \cite{Brown2012} for a perturbative and non-perturbative analysis respectively).   

One may tempted to argue that this effect is due to transient behaviour. This suspicion may become even stronger given that the effect only manifests itself for times comparable to the atomic Heisenberg time. However, what makes our result surprising is that we find no clear evidence that we should associate this behaviour with non-equilibrium transient effects, despite the short interaction time. Rather we find that the response of such detectors can be regarded as non-transient insofar as they satisfy the KMS condition, and a KMS temperature (which decreases with acceleration) can therefore be defined \cite{Kubo1957,Martin1959}. This would mean that these `transients' are of a rather special kind that satisfy detailed balance, a condition which states that each elementary process should be equilibrated by its reverse process, and which is characteristic of equilibrium scenarios. 

\section{Transition probability of an accelerated detector.}  
To model the field-detector interaction it is commonplace to use the  Unruh-DeWitt (UDW) model \cite{DeWitt1979}, 
which consists of a point-like two-level quantum system that couples to a 
scalar field along its trajectory. We will first regard spacetime as a flat static cylinder with spatial circumference $L>0$
(we will later consider the limit $L \rightarrow \infty$).
This cylinder topology is equivalent to imposing periodic boundary conditions relevant to  laboratory systems including closed optical cavities, such as optical-fibre loops \cite{Tsuchida:90},  and superconducting circuits coupled to periodic microwave guides~\cite{Ultrastrong,DCasimir}. 

The coupling of the field to the detector is described by the UDW Hamiltonian \cite{DeWitt1979}
\begin{equation}
H_{I}= \lambda \ \chi(\tau) \mu(\tau) \phi(x(\tau),t(\tau))\label{udw0},
\end{equation}
where $\tau$ is the detector's proper time,  $\mu(\tau)=\sigma_x(\tau)=e^{\ii\Omega \tau}\sigma^++e^{-\ii\Omega \tau}\sigma^-$ is the detector's monopole moment  (with $\sigma^\pm$ being $\mathrm{SU}(2)$ ladder operators), and $\chi(\tau)$ is the switching function. For most of the paper we will consider $\chi(\tau)$ to be Gaussian 
\begin{equation}
\label{swit}\chi(\tau)=e^{-\tau^2/2 \sigma^2}, 
\end{equation}
so that $\sigma$ establishes the timescale of the interaction between the field and the detector. 
The time evolution operator under this Hamiltonian is given by the following perturbative expansion:
\begin{align*}
 \nonumber U=&\openone+U^{(1)}+\mathcal{O}(\lambda^2)=\openone-\ii\int_{-\infty}^{\infty}\!\!\!\text{d} t\, H_{I}(t)+\mathcal{O}(\lambda^2)\\
 =&-\ii\lambda\sum_{m}(I_{+,m}{a}_{m}^{\dagger}\sigma^{+}+I_{-,m}{a}_{m}^{\dagger}\sigma^{-}+\text{H.c.})+\mathcal{O}(\lambda^2),
\end{align*}
where the sum over $m$ takes discrete values due to the periodic boundary conditions ($k=2\pi m/L$).  
$L$ is the scale of the natural IR cutoff (we neglect the interaction of the detector with the zero mode \cite{Jormazero}),
${a_m}$ and $a_m^{\dagger}$ are field mode annihilation and creation operators, and
\begin{equation}\label{Iplus}
I_{\pm,m}= \int_{-\infty}^\infty  \frac{d\tau}{\sqrt{4 \pi |m|}}
	e^{\pm\ii \Omega \tau + \frac{2 \pi  \ii}{L} \left(|m|t(\tau)- m x(\tau)\right)  -\tau^2/2 \sigma^2}, 
\end{equation}
 which can be easily worked out from equation \eqref{udw0}, expanding the field in plane-wave modes and substituting the expression for the monopole moment. If we consider a detector in its ground state, coupled to the vacuum state of the field, the transition probability at leading order in the perturbative expansion, will be given by 
\begin{align}\label{eq:prob1}
\mathcal{P} &\!= \!\!\sum_{m\neq0} | \langle1_m,e | U^{(1)} | 0, g \rangle|^2\!=\!\! \lambda^2\sum_{m\neq 0} | I_{+,m} |^2
\end{align}

\section{Evidence of the `Anti-Unruh' effect.} 
For a uniformly accelerated two-level detector in a periodic cavity, the probability of
transition takes the form 
\begin{equation}\label{eq:prob2}
	\mathcal{P} = \lambda^2 \sum_{ n,\epsilon} \left| \int_{-\infty}^\infty  \frac{d\tau}{\sqrt{4 \pi n}}
	e^{\ii \Omega \tau + 2 \pi n  \ii \left(\frac{ \epsilon}{aL} \left[e^{\epsilon a\tau}-1\right]\right)  -\tau^2/2 \sigma^2} \right|^2
\end{equation}
upon substituting (\ref{Iplus}) into (\ref{eq:prob1}) and using 
\begin{equation} 
\left[|m|t(\tau)- m x(\tau)\right]=\frac{n\epsilon}{a}\left[e^{\epsilon a\tau}-1\right],
\end{equation}
 where $m=-\epsilon n$ where $n \in \mathbb{Z}^+$, $\epsilon=\pm1$,
and $t(\tau)=a^{-1}\sinh(a\tau)$ and $x(\tau)=a^{-1}(\cosh(a\tau)-1)$. As per our comments  in the introduction, when $a\rightarrow0$, $\mathcal{P}$ does not vanish since we are considering a finite time interaction \cite{Louko2008,Brown2012}.

Since the switching function is symmetric about \mbox{$t=0$}, the overall contribution of the right-moving modes is equal to the overall contribution of the left-moving modes, so \eqref{eq:prob2} simplifies to
\begin{equation}\label{eq2}
	\mathcal{P} = 2\lambda^2 \sum_{ n>0} \left| \int_{-\infty}^\infty   \frac{d\tau}{\sqrt{4 \pi n}}
	e^{\ii \Omega \tau + 2 \pi n  \ii \left(\frac{ 1}{aL} \left[e^{ a\tau}-1\right]\right)  -\tau^2/2 \sigma^2} \right|^2,
\end{equation}
which can be recast as 
\begin{align}\label{proba}
	\mathcal{P} \!= \!\frac{-\lambda^2}{2\pi} \int_{-\infty}^\infty\!\!\!\!\!\!  \text{d}\tau \!  \!\int_{-\infty}^\infty\!\!\!\!\!  \text{d}\tau' e^{\ii \Omega (\tau-\tau') -\frac{\tau^2+\tau'^2}{2 \sigma^2}}
	\log[1-e^{ \frac{2 \pi   \ii}{aL} \left[e^{ a\tau}-e^{ a\tau'}\right]} ]
\end{align}
upon summing the series in $n$. The first interesting feature to note in this expression is that the probability is not monotonically increasing with acceleration for all values of the parameters, contrary to expected intuition from the Unruh effect. 

For illustration, before employing the Gaussian switching function, let us first compute the transition rate for sudden switching [which in (1+1) dimensions is finite]. 
Unlike our later results, this rate can be evaluated without requiring high-performance computing.
Consider a detector suddenly switched on at time $t=0$ and switched off at time $t=T$. From (\ref{proba}) (substituting Gaussian by sudden switching) 
the transition rate is
\begin{align}\label{rate}
\dot{\mathcal{P}}=\frac{-\lambda^2}{2\pi}\text{Re}\left(\int_0^T \text{d} s\, e^{\ii\Omega s} \log\left[1-e^{ \frac{2 \pi   \ii}{aL} \left(e^{ aT}-e^{ a(T-s)}\right)} \right]\right)
\end{align}
Plotting this expression as a function of acceleration in Fig. \ref{fig:antiunruhRate} we see that the rate at which this detector clicks can decrease with growing (small) acceleration.
\begin{figure}[htbp]
	\includegraphics[scale=0.65]{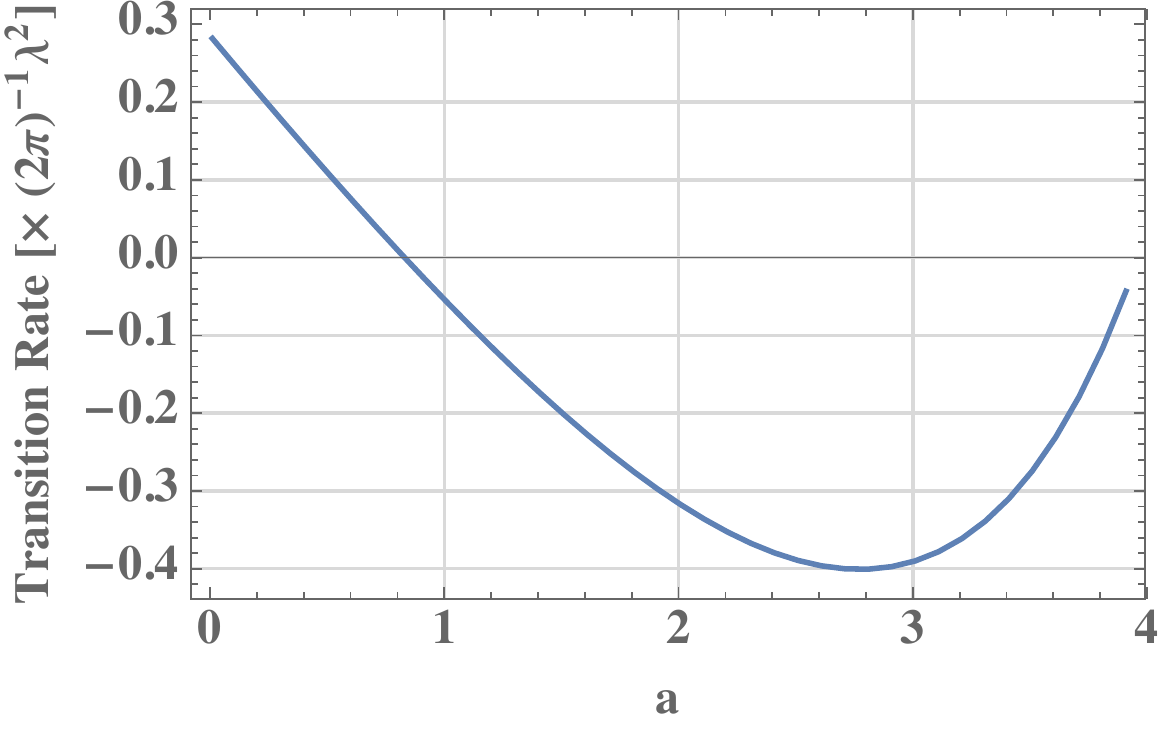}
	\caption{Transition rate (in units of $2\pi\lambda^{-2}$) as a function of acceleration for $T=1,\Omega=2, L=20$. Notice the decreasing transition rate with acceleration for low accelerations.}
	\label{fig:antiunruhRate}
\end{figure}

We find that this phenomenon persists for  Gaussian switching, not only in the transition rate, but also in the transition probability itself. 
However the latter is trickier to evaluate numerically due to  subtleties regarding the singular nature of the 
logarithmically divergent integrand. 
Numerically evaluating  \eqref{eq2} for the Gaussian switching \eqref{swit} we find that the behaviour of the probability is highly dependent on the ratio of the interaction timescale $\sigma$ to the timescale associated with the detector gap $\Omega^{-1}$. Fig. \ref{fig:antiunruh2} displays a plot of \eqref{eq2} for different parameters, showing how varying $\sigma\Omega$ and $L\Omega$ moves us from a regime where the transition probability increases with detector acceleration (as intuitively expected from the Unruh effect), to a regime where this probability decreases with acceleration.
\begin{figure}[htbp]
	\includegraphics[scale=1]{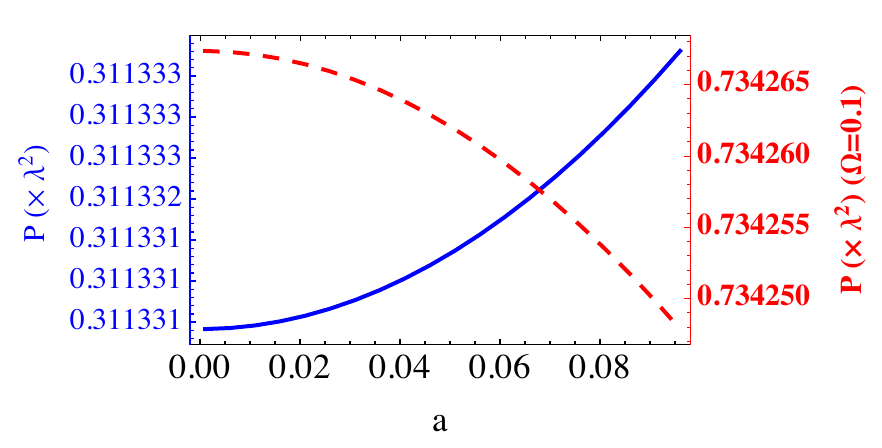}
	\caption{Plot of the transition probability (in units of $\lambda^2$) as a function of acceleration. We set $L=200,\sigma=0.4$ and we vary $\sigma\Omega$ and $L\Omega$ considering both $\Omega= 0.1$ (dashed) and $\Omega= 2$ (solid). For the latter the excitation probability grows with acceleration, whereas for the former it decreases with acceleration. }
	\label{fig:antiunruh2}
\end{figure}

What we have shown so far is that there are regimes for which an accelerated detector in a cavity (for finite times) counts fewer particles as its acceleration grows. 
It is not a sudden switching effect, since it is also present when the switching function is a smooth Gaussian. Could this be due to insufficient interaction time for equilibration? 

To assess this, we will investigate whether or not the detector satisfies the KMS condition \cite{Kubo1957,Martin1959} in this regime. We will find that even though this phenomenon of excitation suppression with increasing acceleration seems to require short times, these times are not so short as to take the system out of the detailed-balance KMS condition. We can therefore use the KMS temperature as a temperature estimator and study how this temperature depends on acceleration for short timescales.

\section{Perturbative analysis of thermality: the KMS condition.} 
Perturbatively, it is commonplace to use the detailed-balance condition obeyed by KMS states \cite{Takagi1986} to evaluate the thermal response of a particle detector. In the context of particle detectors, the KMS condition can be thought of as the postulation that the imbalance between the excitation and de-excitation probabilities of a ground-state and excited detector comes from the equilibrium with a thermal background. To demonstrate thermality, we would need to show a linear dependence of of the logarithm of the
KMS ratio as a function of the gap $\Omega$, where we define the KMS ratio as $\frac{\mathcal{P}(\Omega)}{\mathcal{P}(-\Omega)}$, which for KMS states satisfies
\begin{equation}
	\label{kmsRat}
 \frac{\mathcal{P}(\Omega)}{\mathcal{P}(-\Omega)}=e^{-\Omega/T}.
\end{equation}
\begin{figure}[htbp]
	\includegraphics[scale=0.65]{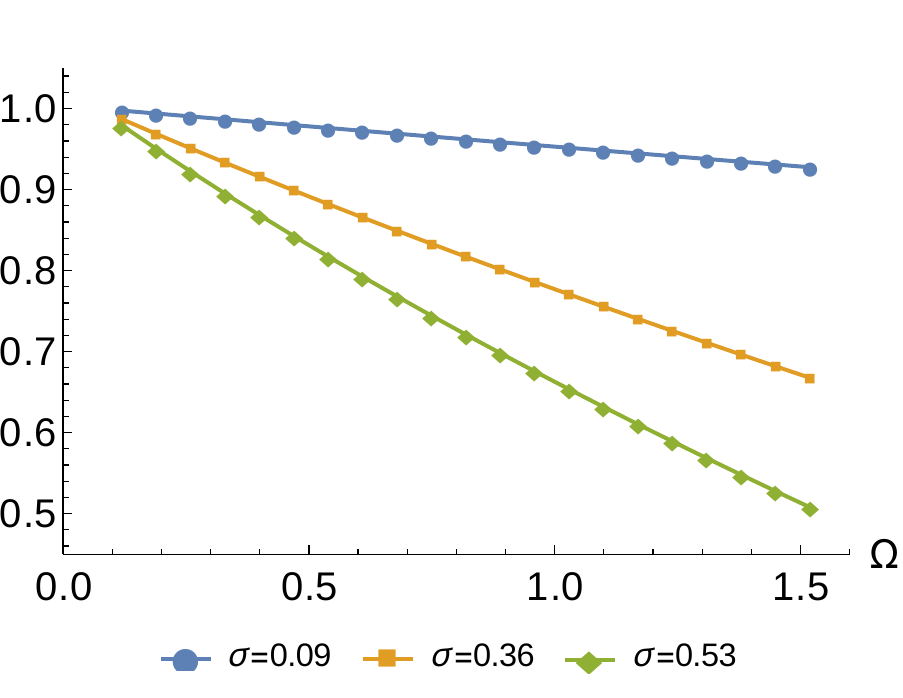}
	\caption{A plot of the logarithm of the KMS ratio versus $\Omega$ for
	$L = 200$, $a=1$. Different lines in the series
	correspond to different values of the Gaussian width, $\sigma$. The relationship is linear even for shorter interaction timescales.}
	\label{fig:kmsratio4}
\end{figure}

This computation was performed in a manner similar to the previous section, though here the use of probability rather than transition rate meant more computing resources were required. The $\mathcal{P}(\Omega)$ was evaluated from equation (\ref{proba}), where the integral was numerically
evaluated over the range [$-10\sigma$, $10\sigma$] so that the error part due to the finite integration is suppressed by $10^{-43}$, well below our numerical
precision. In addition, the number of modes was increased well beyond the point at which the value for the
probability converged within the precision of the temperature.

For given values of $(\sigma,L)$ we computed this KMS ratio for differing values of
$\Omega$;  the temperature was then straightforwardly obtained from equation (\ref{kmsRat}).
A linear slope in the plots of the KMS ratio vs $\Omega$
corresponds to a system that obeys the KMS condition.
Our results are shown in Fig. \ref{fig:kmsratio4}.
We see that the KMS condition is obeyed by the detector for the ranges of parameters considered in the figure.

Consequently we can define a meaningful KMS temperature as the slope of the plot of the KMS ratio as a function of $\Omega$ within this  parameter range. This way we can study the dependence of the KMS temperature 
on the detector's acceleration to identify the regions where the Unruh effect is present. Concretely, we examine the KMS temperature for different values of $\sigma$, $a$, and $\Omega$.

The derivative of the KMS temperature with respect to the acceleration is shown as a density plot in Fig. \ref{fig:antiunruhdensity2}; 
the location where the derivative is zero as a dashed line.
We see that for increasing interaction time (increasing $\sigma$) as well as increasing detector gap $\Omega$,
the negatively sloped region disappears, in line with our expectations that  for long times the slope should approach the usual value of $1/2\pi$. This indicates that turning the detector
on for an infinite amount of time yields the Unruh effect. We also see that the Unruh effect is recovered for large accelerations.

From Fig. \ref{fig:antiunruhdensity2} (top), we see that the temperature change with acceleration increases in magnitude
as acceleration increases. 
Finally, (bottom) we also see that as acceleration increases, the region where the temperature's derivative is negative shrinks,
indicating that we recover the Unruh effect for large accelerations.
 \begin{figure}[htbp]
	\includegraphics[scale=0.65]{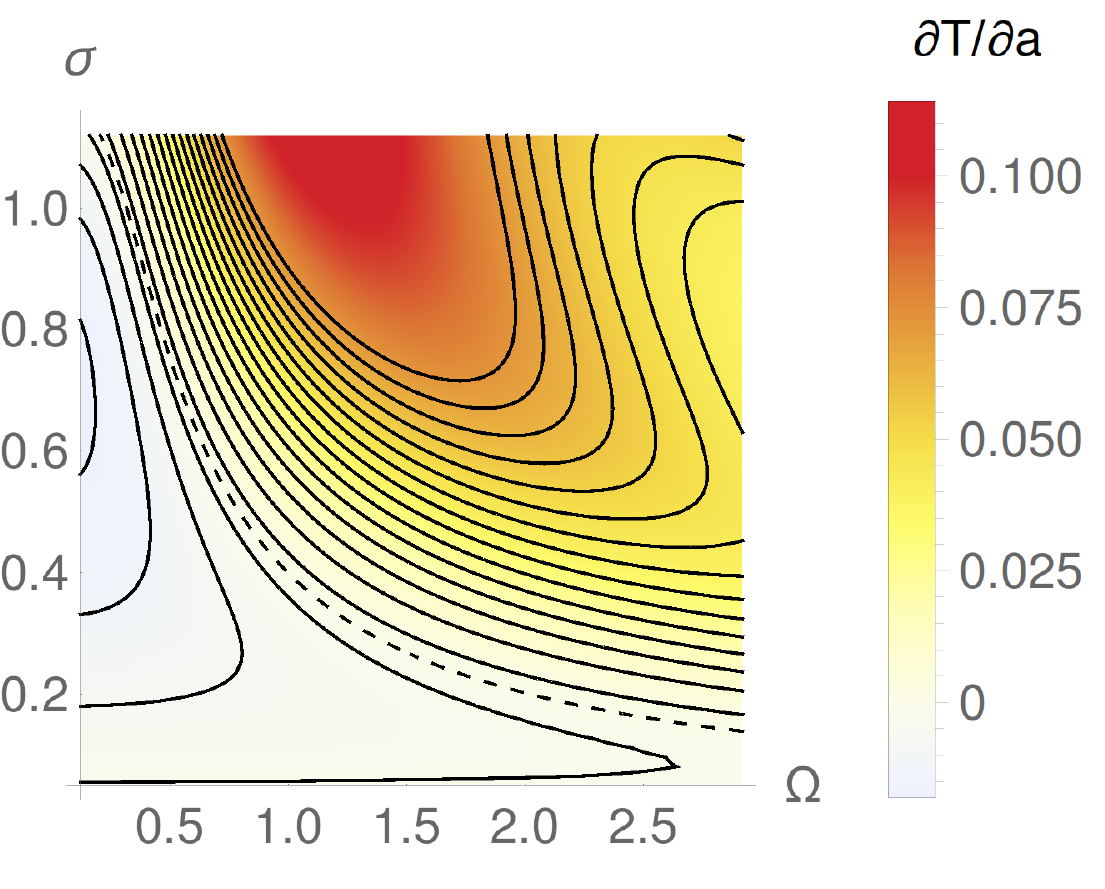}
	\includegraphics[scale=0.65]{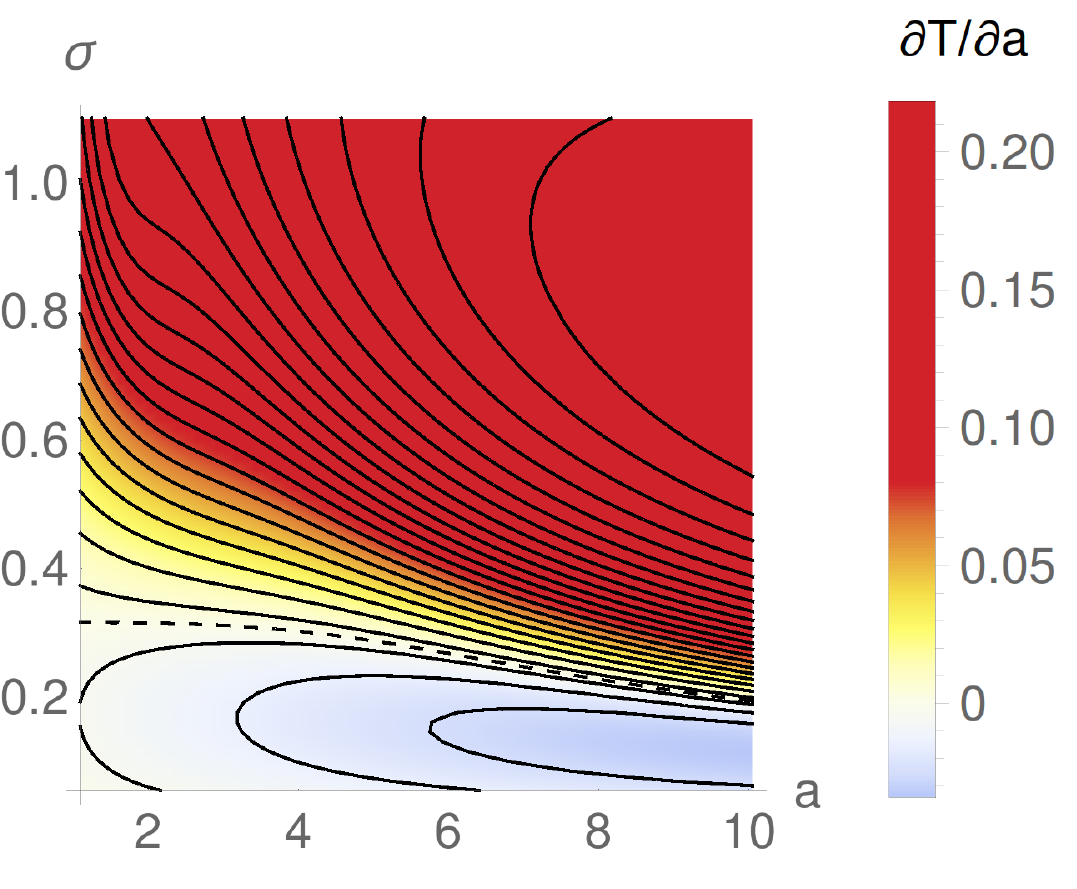}
		\caption{{\bf Top)} Density plot of the $\partial T / \partial a$ versus detector gap $\Omega$ (on the horizontal axis)
	and $\sigma$ (on the vertical axis) for $a = 1.0$. {\bf Bottom)} density plot of the $\partial T / \partial a$ versus acceleration (on the horizontal axis) 	and $\sigma$ (on the vertical axis) for $\Omega = 1.2$. In both plots $L = 200$. }
		\label{fig:antiunruhdensity2}
\end{figure}

\section{(1+1)D continuum case.} 
The effect reported in this letter is not exclusive of cavity setups with periodic boundary conditions. We can examine the effect in the continuum just by replacing the expression \eqref{eq2} by its continuum analogue:
We obtain
\begin{equation}
\label{eqn:intoverk}
\mathcal{P} = \int_{-\infty}^{\infty} \frac{\text{d}k}{4 \pi | k | } 
    \left| \int \text{d}\tau e^{\ii \left[ \Omega \tau  - \frac{s_k}{a} |k| \left( e^{-s_k a \tau} - 1 \right) \right] - \frac{\tau^2}{2 \sigma^2} } \right|^2,
\end{equation}
 where $s_k=\sgn(k)$. We can expand this expression as
\begin{align}
\label{eqn:intoverk2}
\mathcal{P} &= \int_{-\infty}^{\infty} \frac{\text{d}k}{4 \pi | k | }  \int \text{d}t \int \text{d}t'  e^{\ii \Omega (t - t') } e^{- \frac{t^2 + t'^2}{2 \sigma^2}} \times \\
& \quad \quad e^{ - \ii\frac{s_k}{a} |k| \left[ \left( e^{-s_k a t} - 1 \right) - \left( e^{-s_k a t'} - 1 \right) \right]}, \nonumber
\end{align}
a quantity  well known to be IR divergent. Introducing an IR cutoff  $\Lambda$ for regularization, we obtain
\begin{align}
\label{eqn:intoverk3}
\mathcal{P} &= -\int_{-\infty}^{-\Lambda} \frac{\text{d}k}{4 \pi k  }  \int \text{d}t \int \text{d}t'  e^{\ii \Omega (t - t') } e^{- \frac{t^2 + t'^2}{2 \sigma^2}} \nonumber \\
& \quad\quad\quad\qquad \quad \times e^{ - \ii\frac{k}{a}  \left[ \left( e^{ a t} - 1 \right) - \left( e^{ a t'} - 1 \right) \right]} \nonumber\\
&+\int_{\Lambda}^{\infty} \frac{\text{d}k}{4 \pi  k  }  \int \text{d}t \int \text{d}t'  e^{\ii \Omega (t - t') } e^{- \frac{t^2 + t'^2}{2 \sigma^2}}  \nonumber\\
& \quad\quad\qquad \qquad \times e^{ - \ii\frac{k}{a}  \left[ \left( e^{- a t} - 1 \right) - \left( e^{- a t'} - 1 \right) \right]}
\end{align}
which in turn becomes
\begin{align}
\label{eqn:intovert}
\mathcal{P} &= \int_{-\infty}^{\infty} \text{d}t \int_{-\infty}^{\infty} \text{d}t' e^{\ii \Omega \left( t - t' \right)} e^{- \frac{t^2 + t'^2}{2 \sigma^2}} \frac{1}{4 \pi} \\
&  \times \left[ \Gamma_\text{inc}\Big( \frac{\ii \Lambda}{a} ( e^{-a t} - e^{-a t'} )\Big) + \Gamma_\text{inc}\Big( - \frac{\ii \Lambda}{a}  ( e^{a t} - e^{a t'} ) \Big) \right] \nonumber
\end{align}
upon performing the $k$ integral. The analytic continuation of the incomplete gamma function $\Gamma_\text{inc}$ is defined as
\begin{align}
\Gamma_\text{inc}(z)=\int_{z}^{\infty}\! \frac{\text{d}k}{k}e^{-k} 
\end{align}

We can therefore evaluate the expression for the probability of transition for different values of the parameters characterizing the detectors.
\begin{figure}[htbp]
	\includegraphics[scale=1.00]{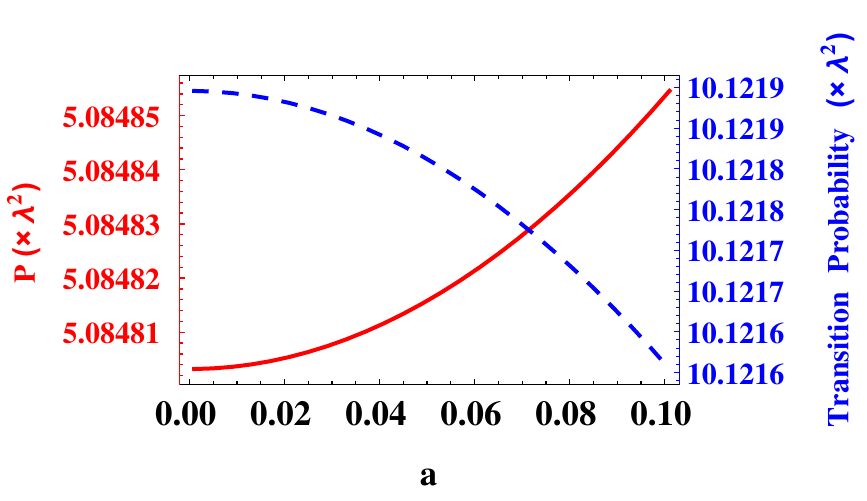}
	\caption{Plot of the transition probability (in units of $\lambda^2$) as a function of acceleration. We set an IR cutoff of $\Lambda = 10^{-7}$, $\sigma=0.8$ for $\Omega= 0.1$ (dashed) and $\Omega= 1$ (solid). For the latter, the excitation probability grows with acceleration, whereas for the former it decreases with acceleration.}
	\label{fig:continuum1}
\end{figure}
The results are depicted in Fig. \ref{fig:continuum1}. We see that as detector acceleration increases, the 
detector can register either more detection events or fewer, depending on the regime of parameter space, demonstrating that this phenomenon is also present in the continuum.

Rather than any kind of boundary conditions,
the key ingredient  responsible for the cooling of an accelerated detector  is the finite time coupling, both for the cavity and the continuum.  Further investigation is required in the latter case to determine if this is  a consequence of the existence of an IR cutoff or the reduced dimensionality of spacetime. 

%

\section{Nonperturbative Thermality.} 
Independent of our study above, we also employed a completely different approach, using a non-perturbative Gaussian formalism  \cite{Brown2012,Brenna2013},  to analyze this phenomenon.  In this scenario the detector is modelled as a harmonic oscillator and ends up in a squeezed thermal state  
upon completion of its interaction with the field; thermality holds 
provided the squeezing contribution to the energy of this state is much smaller than the thermal contribution \cite{Brown2012,Brenna2013}.  We found this criterion to hold for all  values of $(\sigma,\Omega)$ in the relevant parameter regimes of  Fig. \ref{fig:kmsratio4}, 
consistent  with our KMS perturbative analysis: thermality is indeed maintained, even  in the regime where the detector cools  with increasing acceleration. With full disclosure, this nonperturbative calculation was computationally taxing and we were not able to include enough field modes to guarantee full non-perturbative convergence. We therefore can only take this non-perturbative result as an indication, rather than a  non-perturbative proof, of  thermality. We emphasize that our  previous KMS perturbative analysis above is devoid of these limitations.

\section{Discussion of Results.}

We have therefore seen that for certain small-time regimes, the detector
experiences a counterintuitive decrease in temperature given an increase in
acceleration.  The obvious explanation would appear to be that the very short time
scenario induces a non-equilibrium transient effect, which, since it only appears for times of the order of the Heisenberg time of the atom, is not robust, stable, or interesting. However we have found the actual situation to be considerably more subtle for several reasons.

First of all, the effect seems to be stable and robust. We find that the temperature consistently and smoothly decreases with acceleration as we vary all the parameters of the setup.  This indication is further supported by our nonperturbative analysis. Furthermore, the shocking evidence that this is not a typical transient is the apparent thermality of the
effect. To the authors' knowledge, we do not have a better notion for perturbative thermality than the KMS condition, and it indicates that we sustain equilibrium. Either these notions of equilibrium do not
apply for our scenario, or we have an interaction
where the transient effects should not be regarded as non-equilibrium, at least as regards detailed balance.

The magnitude of this effect is very small but this renders it no less interesting. While experimental detection of this effect will challenging (as is detection of the original Unruh effect), we believe that studying the emergence of these phenomena may provide further insight into the relationship between the detailed-balance condition and the thermality of the response of particle detectors in quantum field theory.


We can gain some mathematical insight as to why for small $\sigma\Omega$ the probability of transition can decrease with acceleration and yet one recovers the Unruh effect for longer $\sigma$ by taking a small $\sigma$ expansion of \eqref{eq:prob2}; we can approximately model the salient
features of the detector's transition probability by the following: 
\begin{equation}
	\mathcal{P} \sim \sum_n \frac{\sigma}{n} \left| \int_{-1}^1
	\text{d}\eta\, e^{\ii \sigma \eta \left( \Omega + \frac{2 \pi
	n}{L} \right) } \, e^{\frac{2 \pi \ii n a}{L} \sigma^2 \eta^2} \right|^2
	\label{eqn:analyticProb}
\end{equation}
where the dimensionless parameter $\eta=\tau/\sigma$. There are two different competing trends in \eqref{eqn:analyticProb}. First, for small enough $\Omega$ and $n$, the first exponential in \eqref{eqn:analyticProb} is not highly oscillatory, the second one becomes more oscillatory as the acceleration is increased, therefore the overall value of \eqref{eqn:analyticProb} tends to decrease as acceleration grows. Namely, when $a$ is small, the term $e^{\ii \Omega \sigma  \eta}$ gives the dominant contribution to the integral. In this regime (keeping $n\sim 1$) we see that the integral
reaches a maximum near $\Omega \sigma \sim \pi/2$. As $a$ increases, the integrand becomes more oscillatory and the overall contribution to the integral
decreases. 

Second, to see why the integral increases with acceleration in the Unruh regime, that is, for $\sigma\ll\Omega^{-1}$, we evaluate the integral \eqref{eqn:analyticProb} exactly, and
use the following asymptotic expansion of the imaginary error function
\cite{WolframResearch2001}
\begin{equation*}
	\text{Erfi}(z) \approx \frac{z}{\sqrt{-z^2}} + \frac{1}{\sqrt{\pi} z}
	e^{z^2} \left(1 + \mathcal{O}(1/z^2) \right)
\end{equation*}
for $|z| \rightarrow \infty$.
This essentially (for small $n$) corresponds to a large-$\Omega$ expansion in $\mathcal{P}$ to
$\mathcal{O}(1/\Omega^2)$, and we can actually compute
\begin{equation*}
	\partial_y\!\left| \int_{-1}^1 \!\!\!\!\!e^{\ii x \eta + \ii y \eta^2} \text{d}\eta \right |^2\!\!\!=\!\frac{32 y \left( \left( x^2 + 4 y^2 \right) \cos^2(x) \!+\! 2 x^2 \sin^2(x) \right)}{\left(x^2 - 4 y^2 \right)^3}
\end{equation*}
The expression above is always positive when $x>2y$. From \eqref{eqn:analyticProb} we have
\begin{equation}x=\sigma\left(\Omega+\frac{2\pi n} {L}\right), \qquad y=\frac{2\pi n\, a \sigma^2}{ L}\end{equation}
and so for large $\Omega\sigma$ the probability of excitation will always increase with acceleration provided
\begin{equation}
\Omega>\frac{4\pi n \sigma}{L}a
\end{equation}

The physical intuition that we extract from this analysis is that the vacuum
fluctuations in the short-time regime excite the detector (even if it is
inertial) with a probability that is suppressed by increasing acceleration, more
strongly than the Unruh-like excitation due to the acceleration itself. When the
interaction time is long enough, the vacuum fluctuations get suppressed and the
Unruh effect contribution to the transition probability dominates. While this
indeed suggests features of transient behaviour (short time energy-time
uncertainty), the unexpected feature of the reported phenomenon is that
it preserves detailed balance, and the KMS thermality of the detector.

\section{Conclusions.} 
We have demonstrated that for finite-time interactions $\sigma\sim\Omega^{-1}$, a particle detector at constant acceleration can experience  a cooler heat bath as compared to the same detector with a lower acceleration. Furthermore, when this phenomenon is manifest the KMS condition is satisfied in the same manner as  the usual Unruh effect.  It is quite intriguing that the detailed-balance condition can be satisfied under such circumstances, allowing for a definition of a KMS temperature.  Whether or not it is possible to prolong this effect for longer times remains an interesting open question.

Our results have been restricted to (1+1) dimensions and as such have potential applicability for constrained physical systems such as photons in optical fibres. Extension to (3+1) dimensions may yield further insight into the nature of this effect, and remains a possibility for future work.

\section{Acknowledgments.} We would like to very effusively thank Jorma Louko for extremely inspiring conversations and his always enlightening insight into the physics of accelerated Unruh-DeWitt detectors. 
This work has been supported by the National Sciences and Engineering Research Council of Canada through the Discovery and Vanier CGS programmes.

\bibliography{jabrefaunruh}{}

\end{document}